# Ultra-Thin Absorber based on Phase Change Metamaterial Superlattice

Michael A. Mastro, Virginia D. Wheeler
US Naval Research Lab

**Abstract**
In this paper, a superlattice $VO_2/SiO_2$ metamaterial on a lossy substrate is designed to create a near perfect absorber with tunability across the infrared spectrum. We selected $VO_2$ as it presents a dielectric to metal-like phase change slightly above room temperature. Additionally, the slightly lossy nature of high-temperature $VO_2$ presents comparable and small components (real and imaginary) of the complex refractive index across portions of the visible and infrared. Coupled with a limited conductivity substrate, $VO_2$ has been employed to create highly absorbing/emitting structures where the thickness of the $VO_2$ is ultra-thin (t << $\lambda/4n$). Nevertheless, metal-like $VO_2$ does not possess comparable and small components of the complex refractive index across the entire infrared spectrum, which limits the universality of this ultra-thin $VO_2$ absorber design. Here we employ an ultra-thin superlattice of $VO_2/SiO_2$ to create a composite metamaterial that is readily designed for high absorbance across the infrared spectrum.

## Introduction

A dielectric to metal-like phase change occurs for $VO_2$ at temperatures slightly above room temperature. This has sparked interest in $VO_2$ as a coating on windows of commercial buildings to reflect sunlight on warm days. Unfortunately, this high-temperature reflective state is inherently non-emitting, which is counter to applications such as radiators that need to eject heat at elevated temperatures. This phase change can be generated actively via an applied electric field or a local heating as well as passively via a temperature change of the entire structure or system.

Research at the U.S. Naval Research Lab is partially funded by the Office of Naval Research. U.S. Government work not protected by U.S. copyright.

The use of slightly lossy material was recently proposed to create a single high absorbance layer that is much thinner than the traditional quarter wavelength design. [1] The key insight was to use a material with optical properties intermediate between a (lossless) dielectric and a (perfect electric conductor) metal on a partially lossy substrate. [2] Many semiconductors particularly metal oxides present this intermediate complex dielectric state across large portions of the infrared. The material $VO_2$ is particularly interesting as it undergoes a phase transition near 50ºC. The material is a dielectric below the critical point and is metal-like above the transition point. This is of interest for a variety of applications including switchable antenna structures.

## Design

Previous work on dielectrics with strong optical absorption deposited on metal-like materials with limited conductivity has leveraged the non-trivial reflection phase shift at the interfaces. [2] Specifically, the reflection phase shift is intermediate to 0 and π, when one of the materials has a complex refractive index, $n_c = n + ik$, with comparable n and k. Therefore, the structure can accumulate a near-zero reflectance via multiple optical phase shifts at the interfaces. This is in contrast to the gradual phase change over the optical thickness as in traditional wavelength-scale film optics such as a quarter wavelength mirror.

The criteria to achieve near unity absorbance are a specific interplay of the $VO_2$ thickness, wavelength-dependent $VO_2$ refractive index, and the wavelength-dependent substrate refractive index. Fig 1. displays the absorbance of a $VO_2$ layer as a function of thickness on a sapphire substrate. A high absorbance state is possible for a thickness of 37nm, which is remarkable as this corresponds to an optical thickness of $\lambda/53n$.

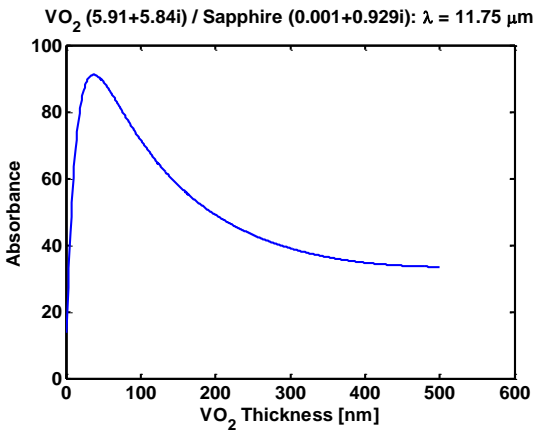

Fig. 1. Absorbance of a VO2 thin film as a function of thickness on a sapphire substrate at a wavelength of 11.75 μm. The peak absorbance occurs for a VO2 film on 37nm.

Although the absorbance is high with a 37nm film, examination of Fig. 2 shows that a significant fraction (approximately 10%) of the infrared light is reflected from this structure. For this particular complex dielectric constant of the $VO_2$ and the sapphire substrate at a wavelength of 11.75 μm, there is not a thickness that allows the reflection coefficients to return to zero.

Inspection of Fig. 3 reveals that a perfect absorbing refractive index condition theoretically exists for a 37nm film on sapphire at a wavelength of 11.75 μm. Unfortunately, this complex refractive index does not match the experimental $VO_2$ refractive index at this wavelength. A summary of Figs 1-3, is that a high absorbance state is available for an ultra-thin $VO_2$ layer on sapphire at this wavelength, but a perfect absorber/emitter is not available at any thickness.

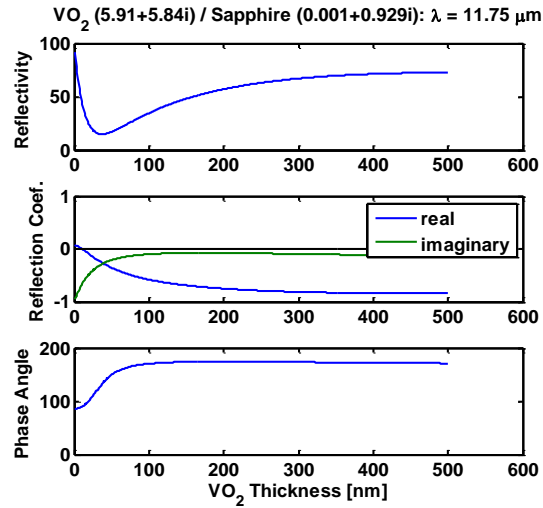

Fig. 2 (top) Reflectivity, (middle) reflection coefficient, and (bottom) phase angle of a $VO_2$ thin film as a function of thickness on a sapphire substrate at a wavelength of 11.75 μm.

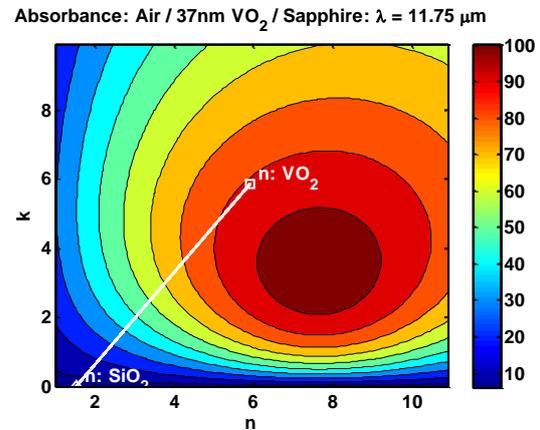

Fig. 3. Theoretical absorbance of a 37nm VO2 layer on sapphire at a wavelength of 11.75 μm as a function of real (n) and imaginary (k) complex refractive index coefficient. A near unity absorbance condition exists for a hypothetical film with a refractive index n = 8, and k = 4. For reference, the experimental refractive index of $VO_2$ and $SiO_2$ are also displayed.

A solution to this non-unity absorber limitation, is to form a superlattice with another layer. Fig. 4 show that 2nm $VO_2$ / 8 nm $SiO_2$ superlattice displays an absorbance of approximately 92%. As each individual $VO_2$ and $SiO_2$ layer thickness is much less than the infrared wavelength, it is safe to describe the superlative as a composite layer.

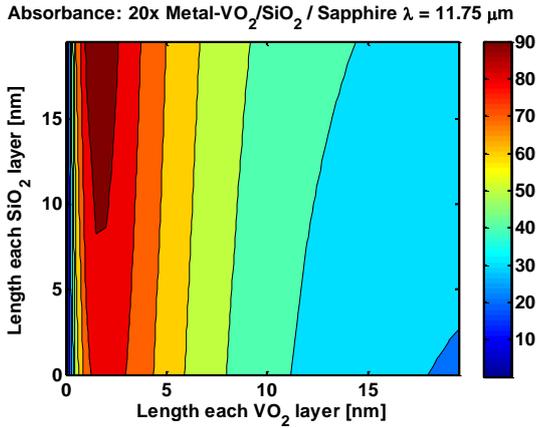

Fig. 4. Absorbance of a 20x VO₂/SiO₂ superlattice as a function of VO₂ and SiO₂ layer thickness.

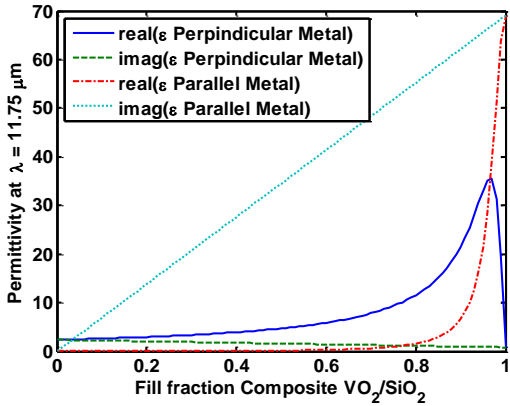

Fig. 5. Effective parallel and perpendicular permittivity of the VO₂/SiO₂ superlattice as a function of fill fraction. The parallel and perpendicular terms are relative to the surface of the superlative.

Viewing the superlattice as a single composite layer provides greater insight into the behavior of the material. In the superlattice the effective perpendicular permittivity can be expressed by

$$\varepsilon_\perp = \frac{\varepsilon_{Metal}\varepsilon_{Dielectric}}{f\varepsilon_{Metal}+(1-f)\varepsilon_{Dielectric}} = \frac{1}{\frac{d_{Metal}/\varepsilon_{Metal}+d_{Dielectric}/\varepsilon_{Dielectric}}{d_{Metal}+d_{Dielectric}}},$$
(1)

where f is the fill fraction of the metallic VO₂. Additionally, in the superlattice the parallel permittivity can be expressed by

$$\varepsilon_\parallel = f\varepsilon_{Metal}+(1-f)\varepsilon_{Dielectric} = \frac{d_{Metal}\varepsilon_{Metal}+d_{Dielectric}\varepsilon_{Dielectric}}{d_{Metal}+d_{Dielectric}},$$
(2)

where the metal term refers to the metallic VO₂ and the dielectric term refers to the SiO₂.

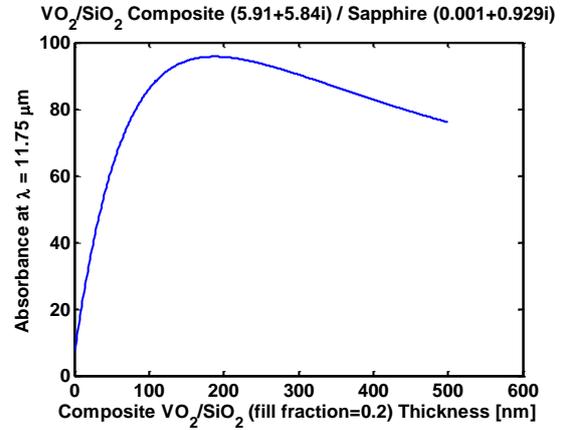

Fig. 6. Absorbance of a VO₂-SiO₂ composite layer as a function of thickness on a sapphire substrate. The calculation is based on an effective parallel complex refractive index of 2.8295 + 2.4399i interacting with an on-axis incident wave.

The absorbance of this composite stack as a function of thickness on a sapphire substrate is observable in Fig. 6. The peak absorbance of 97% for a thickness of 190nm is remarkable in that the total optical thickness is λ/22n. Creating a sub-wavelength superlattice metamaterial allow the effective creation of new optical properties that allow near unity absorption.

Conclusion

The paper present a simple technique to ease the application and increase effectiveness of ultra-thin lossy layers as the basis for near-unity absorber/emitter structures. Additionally, the phase change nature of VO₂ is useful for systems and devices that require a passive or active change in emittance.

**Acknowledgments**

The work at NRL was partially supported by the Office of Naval Research.